\documentclass[cits]{PoS}

\title{Baryon spectrum with Nf=2+1+1 twisted mass fermions}

\ShortTitle{Baryon spectrum with Nf=2+1+1 twisted mass fermions}

\author{Constantia Alexandrou\\
        Department of Physics, University of Cyprus, P.O. Box 20537, 1678 Nicosia, Cyprus and \\
        Computation-based Science and Technology Research Center, The Cyprus Institute, 20 Kavafi Str., Nicosia 2121, Cyprus\\
        E-mail: \email{alexand@ucy.ac.cy}}
        
\author{Vincent Drach\\
        CP$^3$-Origins and the Danish Institute for Advanced Study DIAS, University of Southern Denmark, Campusvej 55, DK-5230 Odense M, Denmark\\
        E-mail: \email{drach@cp3.dias.sdu.dk}}

\author{Kyriakos Hadjiyiannakou\\
       Department of Physics, University of Cyprus, P.O. Box 20537, 1678 Nicosia, Cyprus\\
        E-mail: \email{hadjigiannakou.kyriakos@ucy.ac.cy}}

\author{Karl Jansen\\
        NIC, DESY, Platanenallee 6, D-15738 Zeuthen, Germany\\
        E-mail: \email{Karl.Jansen@desy.de}}

\author{\speaker{Christos Kallidonis}\\
       Computation-based Science and Technology Research Center, The Cyprus Institute, 20 Kavafi Str., Nicosia 2121, Cyprus\\
       E-mail: \email{c.kallidonis@cyi.ac.cy}}

\author{Giannis Koutsou\\
        Computation-based Science and Technology Research Center, The Cyprus Institute, 20 Kavafi Str., Nicosia 2121, Cyprus\\
        E-mail: \email{g.koutsou@cyi.ac.cy}}

\abstract{We present results on the masses of the low-lying baryons using ten ensembles of gauge configurations with $N_f=2+1+1$ dynamical twisted mass fermions, at three values of the lattice spacing, spanning a pion mass range from about 210 MeV to about 430 MeV. The strange and charm quark masses are tuned to approximately their physical values. We examine isospin symmetry breaking effects on the baryon mass and the dependence on the lattice spacing. After taking the continuum limit we use  chiral perturbation theory to extrapolate to the physical vlaue of the pion mass for all forty baryons. We provide predictions for the masses of doubly and triply charmed baryons that have not yet been measured experimentally.}

\vtop{\hbox{\hspace{7.28cm}CP3-Origins-2014-035 DNRF90, DIAS-2014-35}}

\FullConference{The 32nd International Symposium on Lattice Field Theory,\\
		23-28 June, 2014\\
		Columbia University New York, NY}

\def\fig#1{Fig. \ref{#1}}

\begin{document}

\section{Introduction}

The remarkable progress that has been achieved in lattice QCD (LQCD) during the last years allows for simulations using light quark masses closer to their physical values. This leads in more accurate chiral extrapolations to the physical pion mass and thus a reliable calculation of the low-lying hadron masses.
This work focuses on a LQCD study of the masses of the forty low-lying hyperons and charmed baryons using $N_f=2+1+1$ dynamical twisted mass fermions at maximal twist, which ensures an $\mathcal{O}(a^2)$ behavior of our results.
A total of ten ensembles are analyzed at three values of the lattice spacing, enabling us to take the continuum limit and perform chiral extrapolations to the physical pion mass. The good precision of our results allows for comparisons with experiment and reliable predictions for the masses of doubly and triply charmed $\Xi$ and $\Omega$ baryons.


\section{Setting the scale}

\begin{figure}[!ht]\vspace*{-0.2cm}
\begin{minipage}{8cm}
\includegraphics[width=0.9\textwidth]{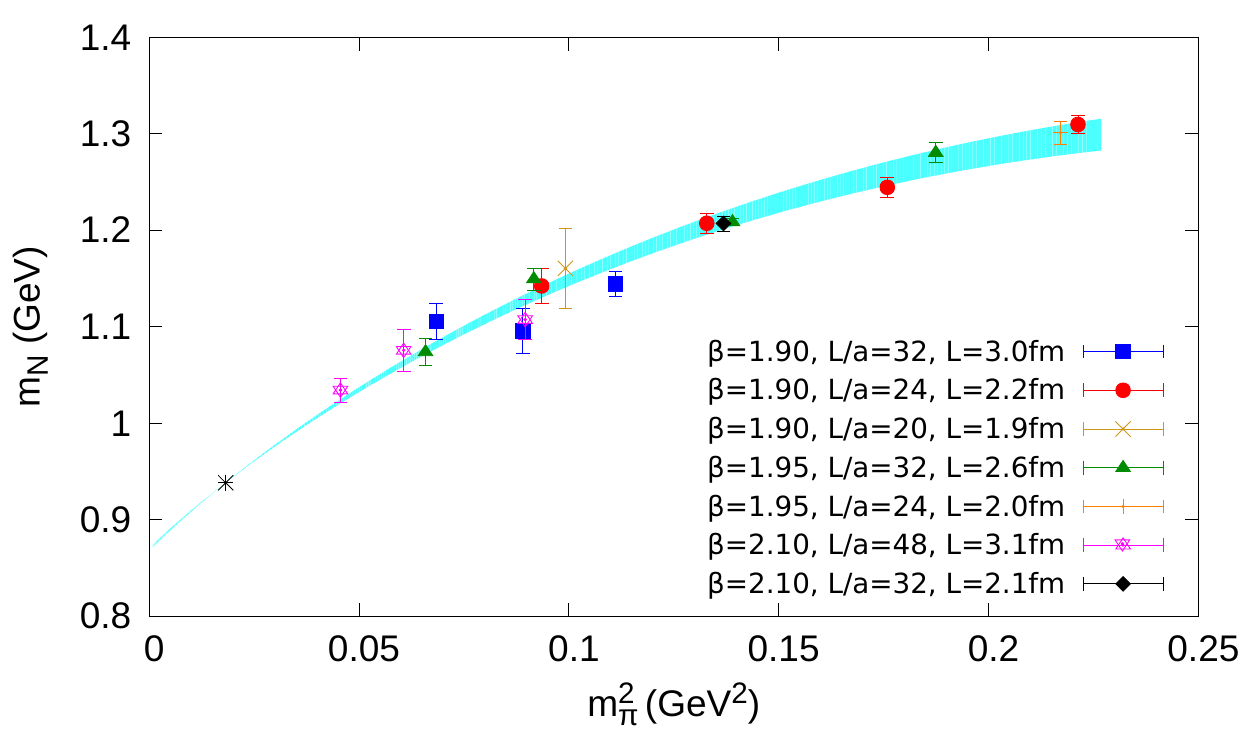}
\end{minipage}\hfill
\begin{minipage}{8cm}
\includegraphics[width=0.9\textwidth]{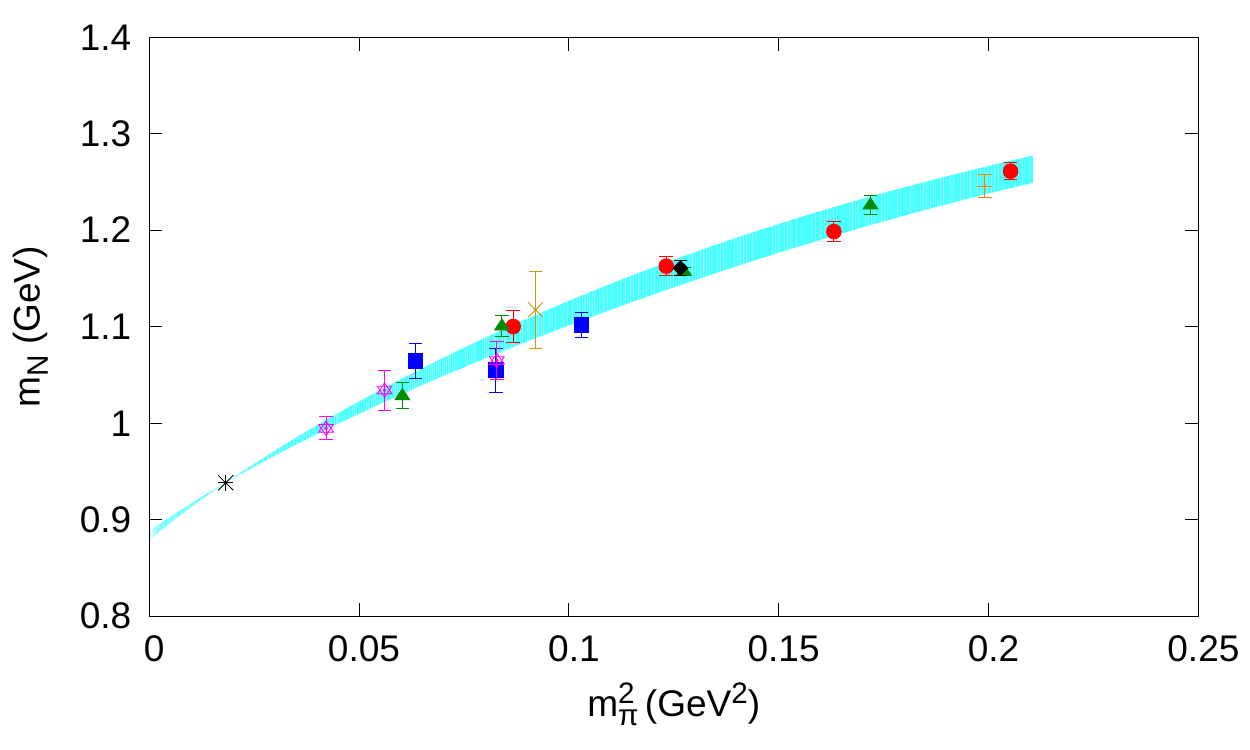}
\end{minipage}
\caption{Fit to the nucleon mass treating the lattice spacings as fit parameters. The physical nucleon mass is denoted with the asterisk. Left: The fit to $\mathcal{O}(p^3)$ expansion from HB$\chi$PT. Right: The fit to $\mathcal{O}(p^4)$ with explicit $\Delta$ degrees of freedom. The notation is given in the legend of the left plot.}
\label{fig:nucleon_mass}
\end{figure}

When calculating baryon masses, the physical nucleon mass is an appropriate quantity to set the scale. To this end we carried out a high statistics analysis of the nucleon mass on a total of 17 $N_f=2+1+1$ gauge ensembles in order to obtain an accurate determintion of the lattice spacings. For the chiral fits we used the well established result from HB$\chi$PT, $m_N=m_N^{(0)} -4c_1m_\pi^2 - \frac{3g_A^2m_\pi^3}{16\pi f_\pi^2}$. Assuming no cut-off effects in the case of the nucleon, we fitted simultaneously for all $\beta$ values, treating the lattice spacings $a_{\beta=1.90}$, $a_{\beta=1.95}$ and $a_{\beta=2.10}$ as additional fit parameters. We estimate a systematic error due to the chiral extrapolation by performing the fit using an $\mathcal{O}(p^4)$ expression from HB$\chi$PT with explicit $\Delta$-degrees of freedom. The resulting fits are shown in \fig{fig:nucleon_mass}. Both expressions describe well our lattice data and they fall on a universal curve. The three values of the lattice spacing obtained from this combined fit are $a_{\beta=1.90}=0.0936(13)(35)$~fm, $a_{\beta=1.95}=0.0823(10)(35)$~fm and $a_{\beta=2.10}=0.0646(7)(25)$~fm, where the error in the first parenthesis is the statistical and in the second parentehesis the systematic due to the chiral extrapolation, estimated by taking the difference between the values obtained in the two fits. 
 Performing the fit separately for each $\beta$ value yields values for the lattice spacing which are consistent with those obtained from the combined fit, indicating that indeed cut-off effects are negligible in the nucleon case. According to the Hellman-Feynamn theorem, the $\mathcal{O}(p^3)$ and $\mathcal{O}(p^4)$ expressions for the nucleon mass   can be used to provide an estimate of the light $\sigma$-term of the nucleon, $\sigma_{\pi N}$. From our fits we find $\sigma_{\pi N} = 64.9(1.5)(19.6)$~MeV.


\section{Tuning of the strange and charm quark masses}

In order to avoid complications due to flavor mixing in the heavy quark sector, we employ the Osterwalder-Seiler setup of valence strange and charm quarks. This requires a tuning of the bare strange and charm quark masses. Since we are interested in baryon spectrum, we choose to use the physical mass of the $\Omega^-$ ($\Lambda_c^+$) baryon to fix the strange (charm) quark mass to approximately its physical value. For the tuning we used the non-perturbatively determined renormalization constants $Z_P$ computed in Ref.~\cite{Carrasco:2014cwa} in the $\overline{\rm MS}$ scheme at 2 GeV.

\begin{figure}[!ht]\vspace*{-0.2cm}
\begin{minipage}{7cm}
\includegraphics[width=\textwidth]{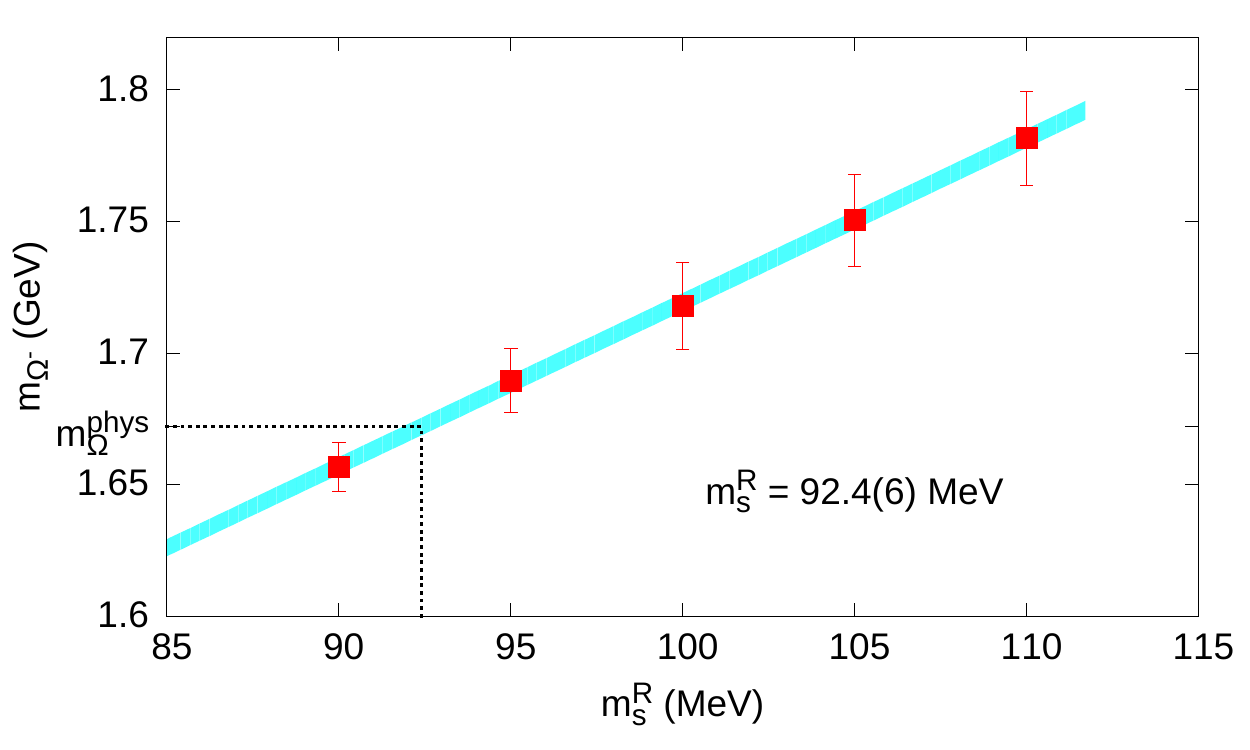}
\end{minipage}\hfill
\begin{minipage}{7cm}
\includegraphics[width=\textwidth]{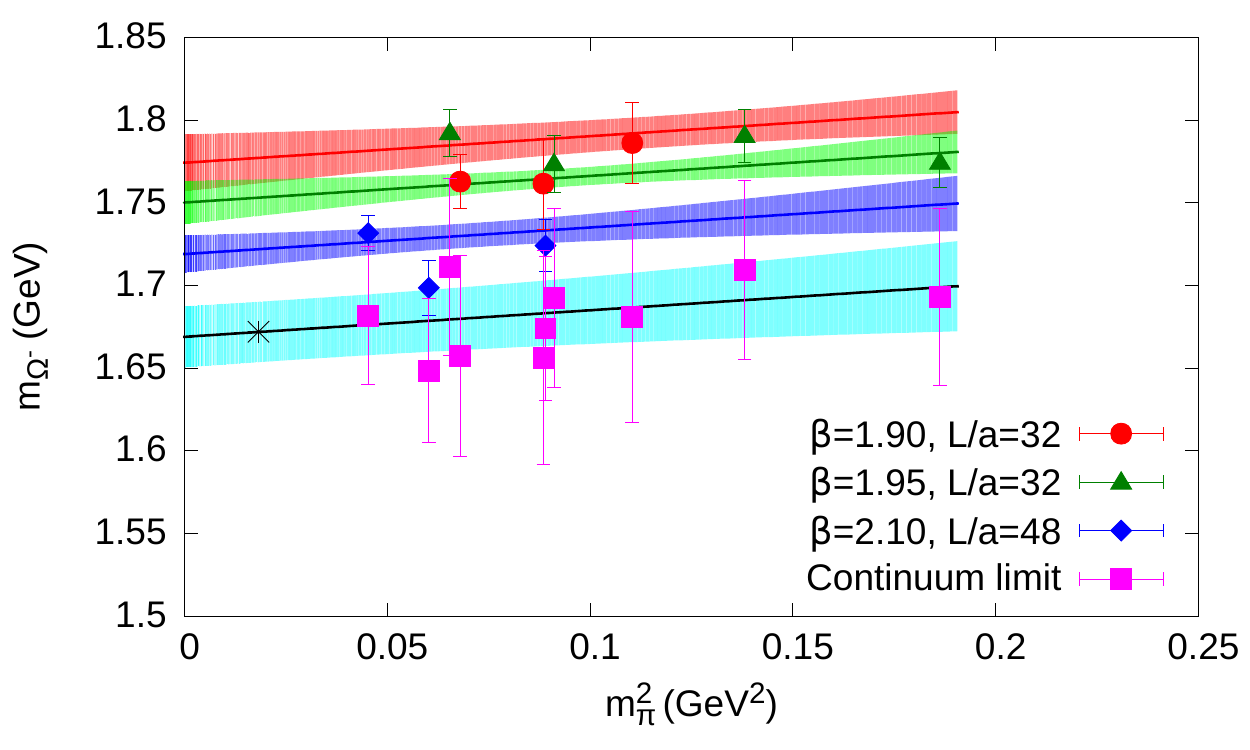}
\end{minipage}
\caption{Left: Matching of the $m_s^R$ with the physical $\Omega^-$ mass. Right: Chiral and continuum extrapolation of the $\Omega^-$ mass at the tuned value of $m_s^R$. The physical $\Omega^-$ mass is reproduced.}
\label{fig:omega_tuning}
\end{figure}

The concept was to use several values of the bare strange and charm quark mass to interpolate the $\Omega^-$ ($\Lambda_c^+$) mass to certain values of the renormalized strange (charm) quark mass and then extrapolate to the physical pion mass and the continuum. During the process, the value of the renormalized quark mass is changed iteratively until the extrapolated baryon mass agrees with the experimental one. This determines the tuned value of $m_s^R$ and $m_c^R$ that reproduces the physical mass of $\Omega^-$ and $\Lambda_c^+$, respectively. For the extrapolations we used the expressions $m_\Omega = m_\Omega^{(0)} - 4c_\Omega^{(1)}m_\pi^2$ from SU(2) $\chi$PT and $m_{\Lambda_c} = m_{\Lambda_c}^{(0)} + c_1m_\pi^2 + c_2m_\pi^3$, motivated by SU(2) HB$\chi$PT. Cut-off effects are taken into account by adding a quadratic term $da^2$ to these expressions, where $d$ is an additional fit parameter. The values we find by following this procedure are $m_s^R=92.4(6)(2.0)$~MeV and $m_c^R=1173.0(2.4)(17.0)$~MeV. The error in the first parenthesis is the statistical and in the second parenthesis is the systematic, calculated by allowing the renormalized mass to vary within the statistical errors of the $\Omega^-$ and $\Lambda_c^+$ at the physical pion mass. In \fig{fig:omega_tuning} we show representative plots for  tuning the strange quark mass.
%

\section{Results I: Isospin symmetry breaking}

\begin{figure}[!ht]\vspace*{-0.2cm}
\begin{minipage}{5cm}
\includegraphics[width=\textwidth]{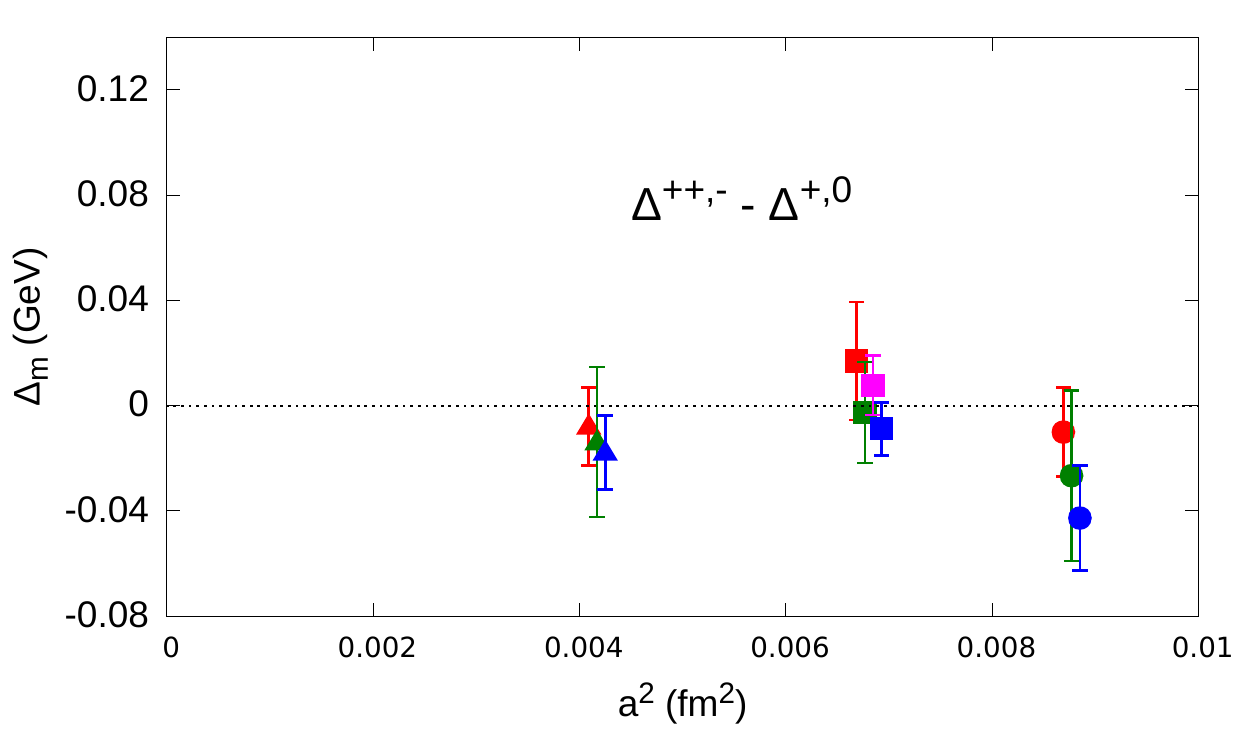}
\end{minipage}\hfill
\begin{minipage}{5cm}
\includegraphics[width=\textwidth]{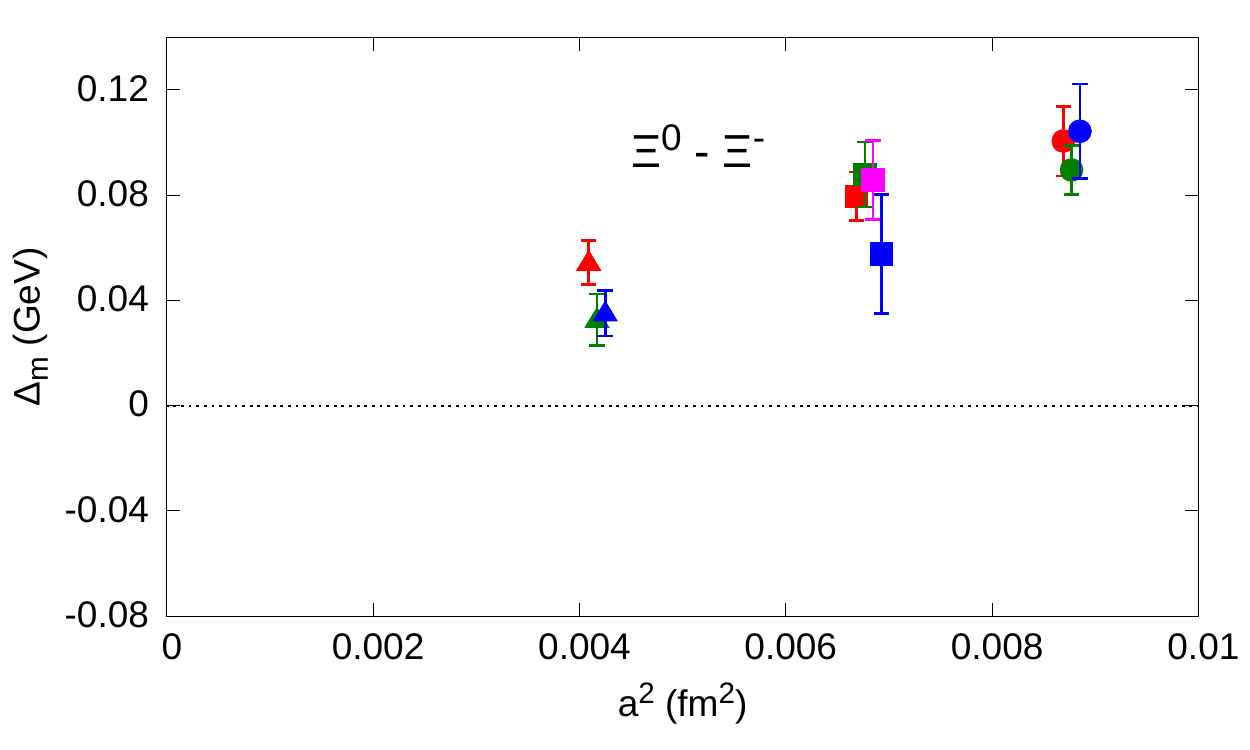}
\end{minipage}
\begin{minipage}{5cm}
\includegraphics[width=\textwidth]{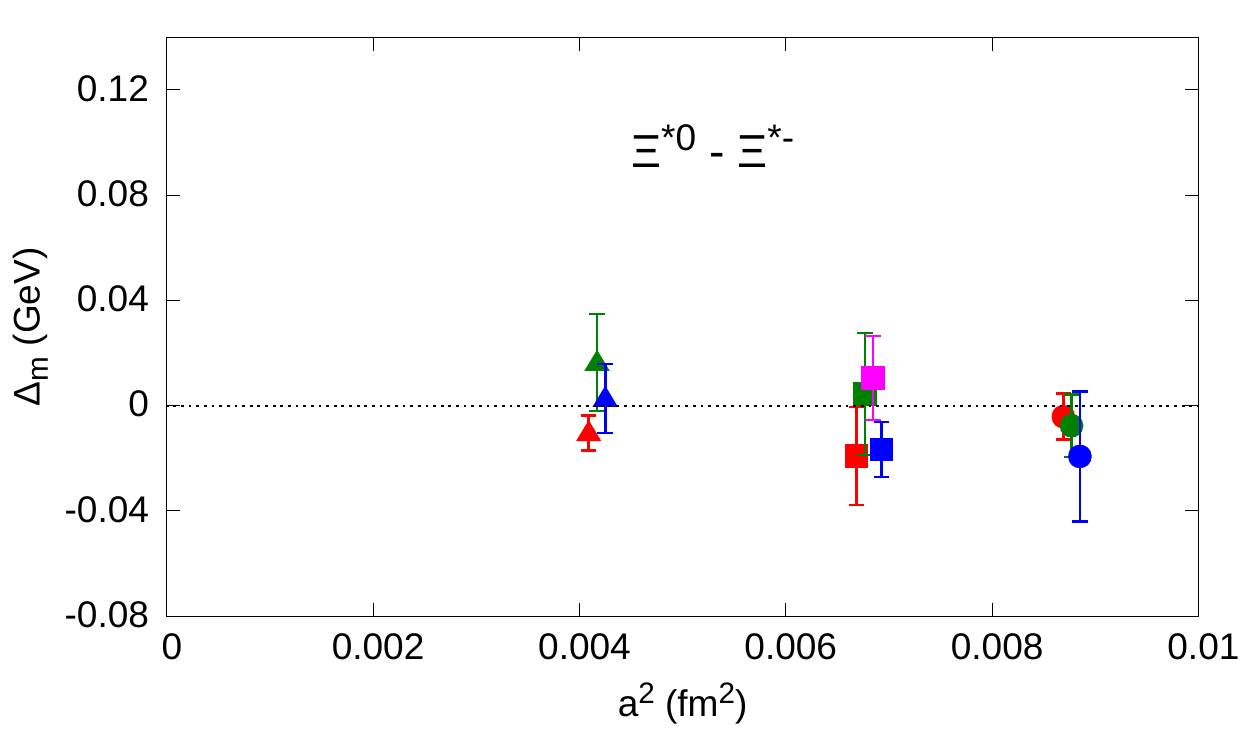}
\end{minipage}
\begin{minipage}{5cm}
\includegraphics[width=\textwidth]{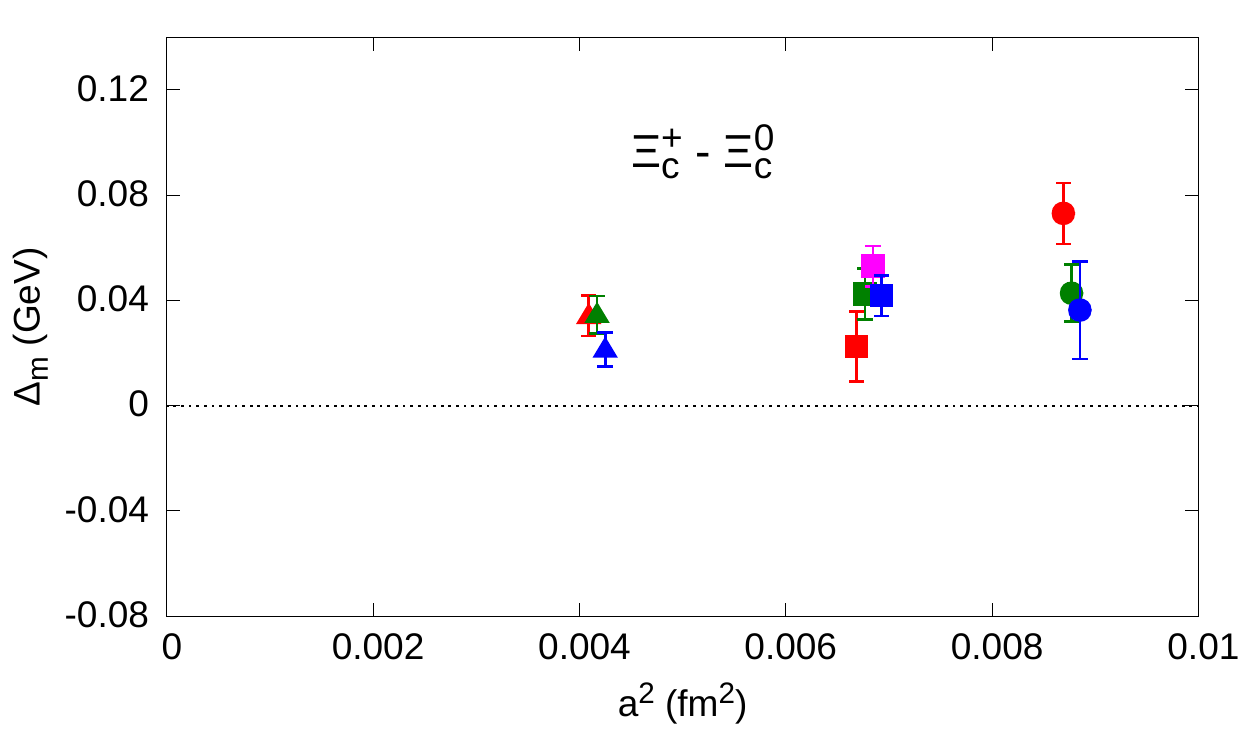}
\end{minipage}\hfill
\begin{minipage}{5cm}
\includegraphics[width=\textwidth]{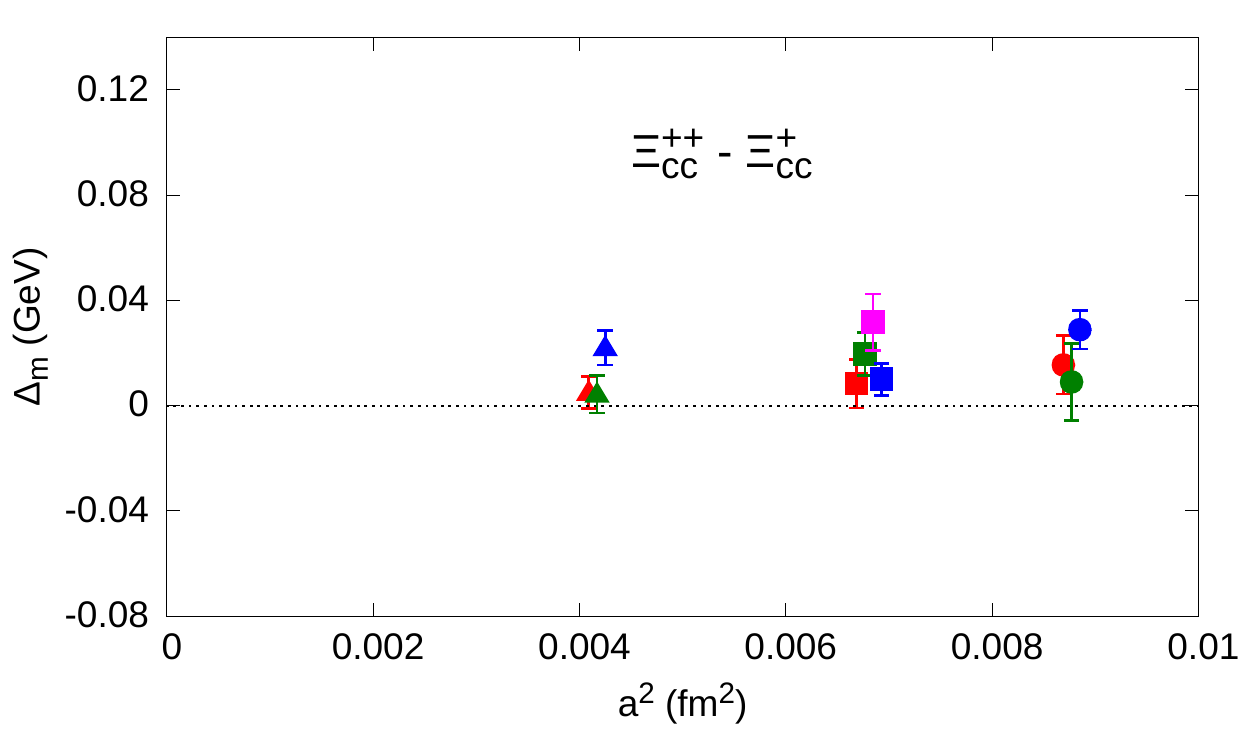}
\end{minipage}\hfill
\begin{minipage}{5cm}
\includegraphics[width=\textwidth]{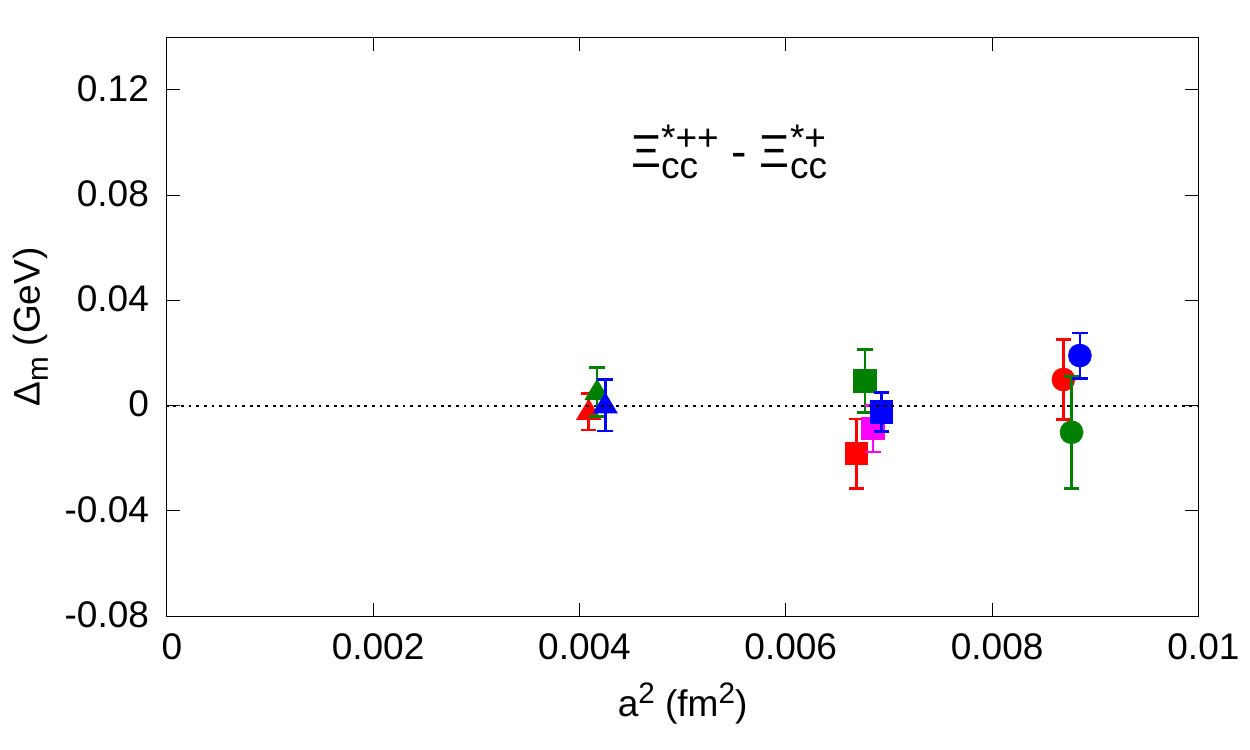}
\end{minipage}
\caption{Top: Mass difference for the $\Delta$ baryons (left), the spin-1/2 $\Xi$ baryons (center) and the spin-3/2 $\Xi^*$ baryons (right). Bottom: Mass difference for the $\Xi_c$ baryons (left), the spin-1/2 $\Xi_{cc}$ baryons (center) and the spin-3/2 $\Xi_{cc}^*$ baryons (right)}
\label{fig:isospin}
\end{figure}

The Wilson twisted mass action which we employ in our calculations breaks isospin symmetry explicitly to $\mathcal{O}(a^2)$. It manifests itself as mass splitting between baryons belonging to the same isospin multiplets. 
We show representative plots of the mass difference as a function of $a^2$ for a number of octet and decuplet isospin multiplets in the top panel of \fig{fig:isospin}. As can be seen for the $\Delta$ baryons, the mass difference is consistent with zero indicating that isospin breaking effects are small for the $\beta$ values analysed. For the spin-1/2 hyperons we observe small differences which decrease linearly with $a^2$ being almost zero at our smallest lattice spacing, while the mass difference for the spin-3/2 baryons is consistent with zero at all lattice spacings. This is shown by the mass differences for $\Xi$ and $\Xi^*$ in \fig{fig:isospin}. Extending this analysis for the charm baryons we plot the mass differences for the $\Xi_c$ multiplet as well as the doubly charmed $\Xi_{cc}$ baryons in the bottom panel of \fig{fig:isospin}. While small non-zero differences exist for the $\Xi_c$ case, one can see that isospin splitting is consistent with zero at all lattice spacings for both the spin-1/2 and spin-3/2 doubly charmed $\Xi_{cc}$ baryons. 

Additionally, for a given lattice spacing one can examine the dependence of the isospin mass splitting on the pion mass. As shown in \fig{fig:isospin} the baryon mass differences do not depend on the light quark mass within our statistical accuracy.


\section{Results II: Chiral extrapolations}

When extrapolating our lattice results to the physical pion mass we allow for cut-off effects by including a term $da^2$ where $d$ is treated as a fit parameter, and we then apply continuum chiral perturbation theory at our results. The fits are performed in the whole pion mass range of about 210-430 MeV and all $\beta$ values are included.

\begin{figure}[!ht]\vspace*{-0.2cm}
\begin{minipage}{7cm}
\includegraphics[width=\textwidth]{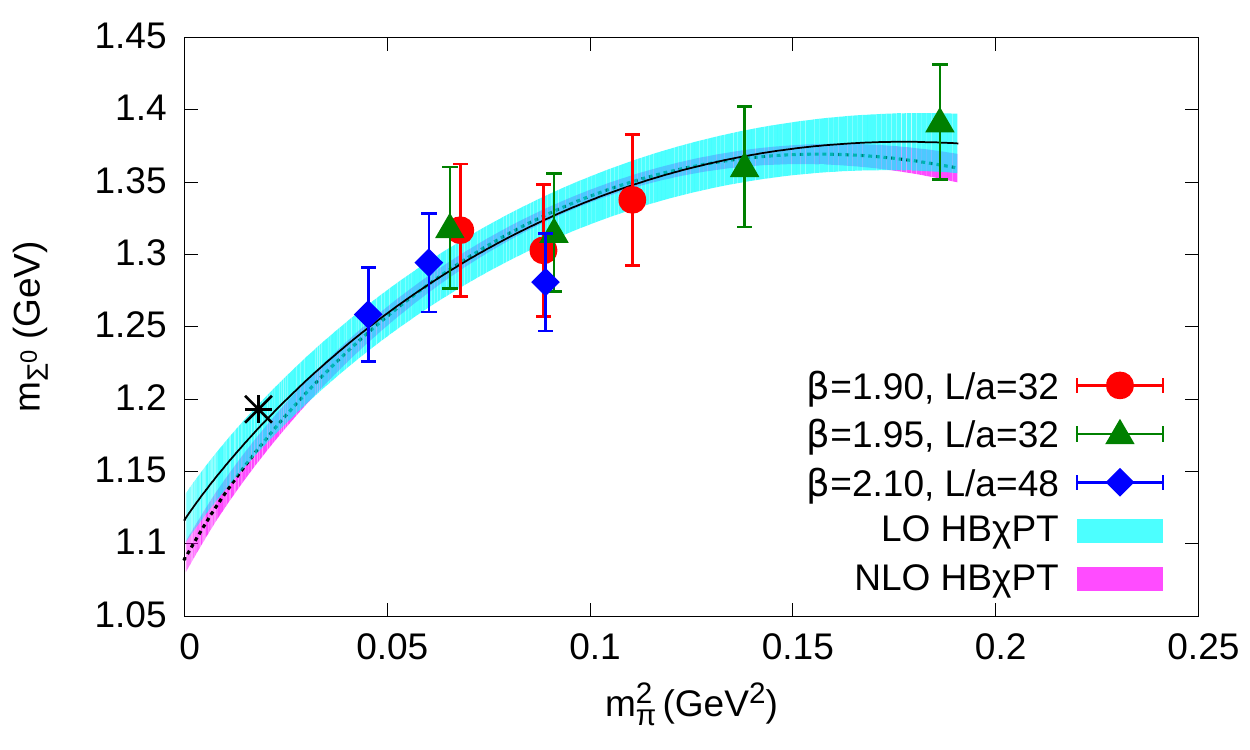}
\end{minipage}\hfill
\begin{minipage}{7cm}
\includegraphics[width=\textwidth]{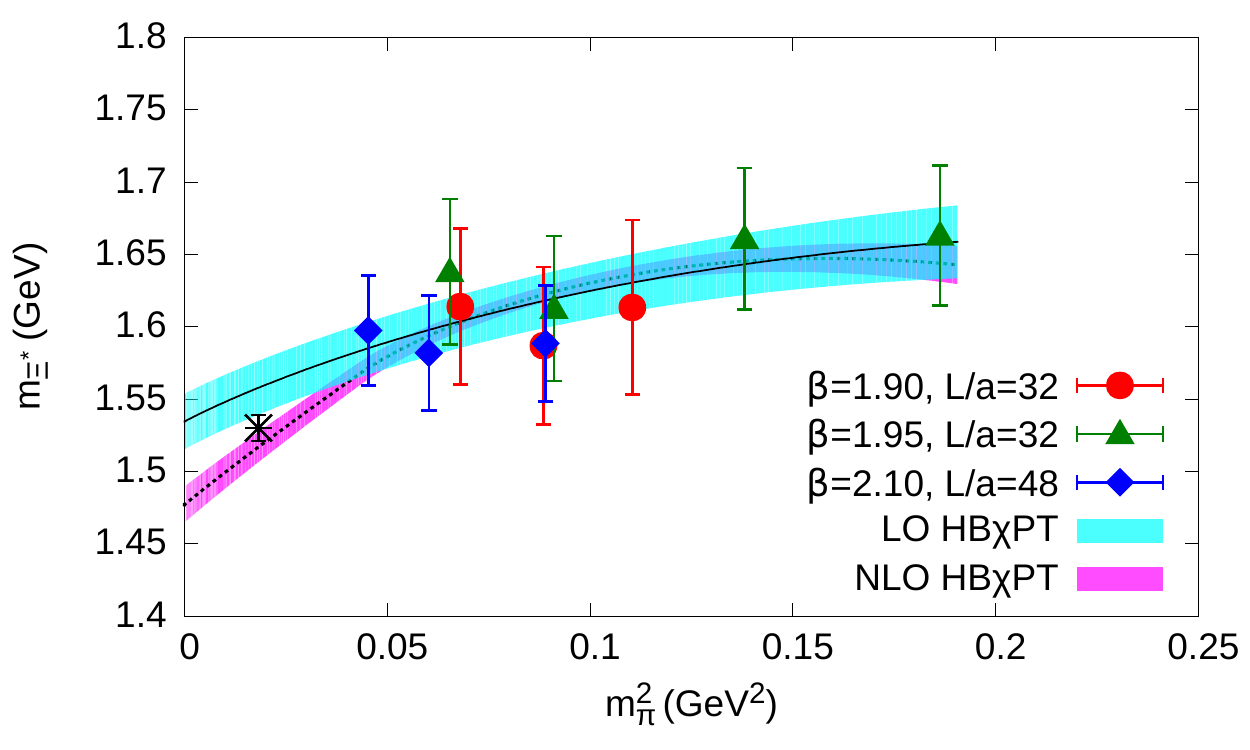}
\end{minipage}
\vspace{-0.4cm}
\begin{minipage}{7cm}
\includegraphics[width=\textwidth]{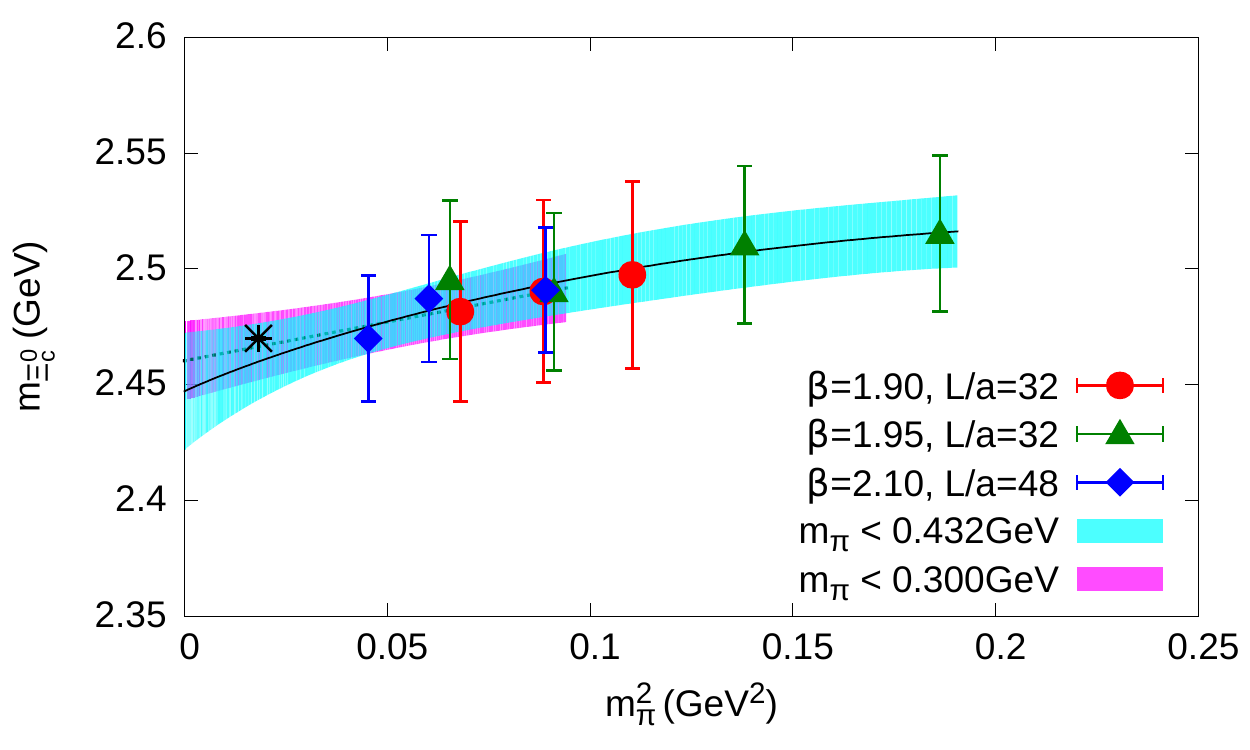}
\end{minipage}\hfill
\begin{minipage}{7cm}
\includegraphics[width=\textwidth]{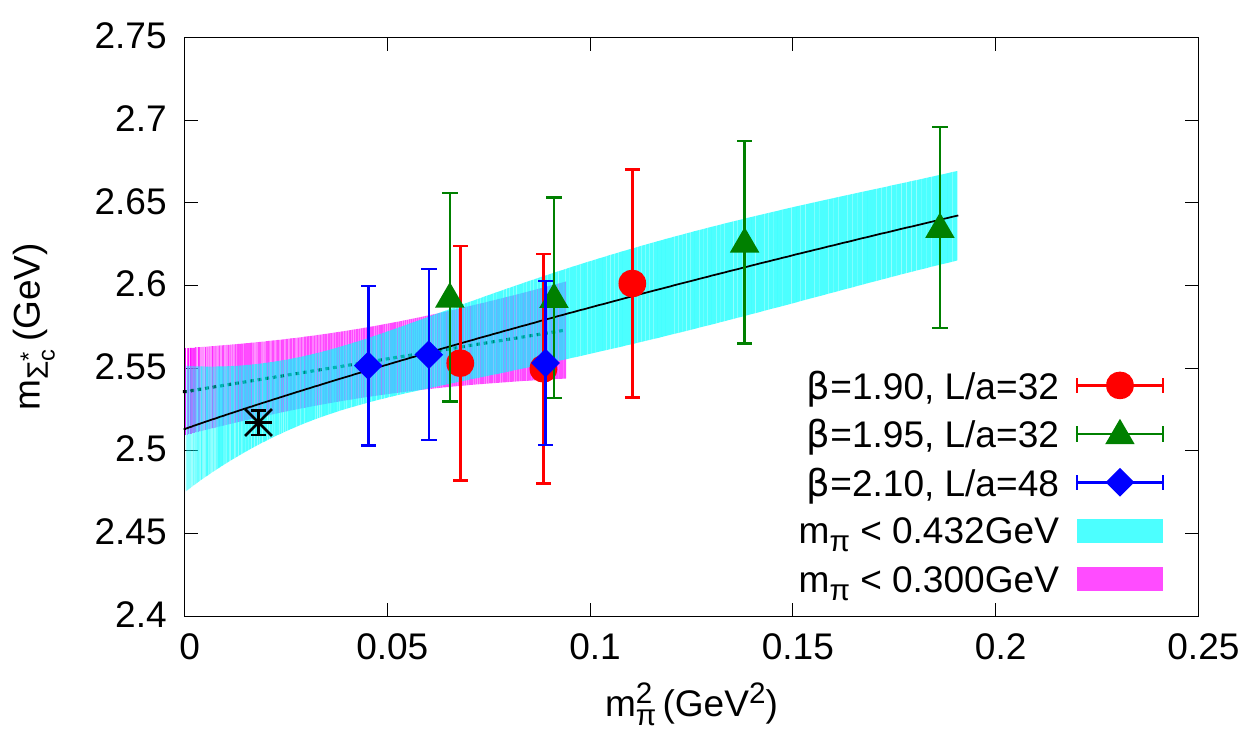}
\end{minipage}
\caption{Top: Chiral extrapolations for $\Sigma^0$ (left) and $\Xi^*$ (right) in physical units using the LO and NLO expressions from $\chi$PT. Bottom: Chiral extrapolations for $\Xi_c^0$ (left) and $\Sigma_c^*$ (right) in physical units using the Ansatz as explained in the text. The notation is given in the legends of the plots. The experimental value is shown with the black asterisk.}
\label{fig:chiral}
\end{figure}

For the octet and decuplet baryons, we consider the leading one-loop expressions from SU(2) HB$\chi$PT~\cite{Nagels:1978sc,Nagels:1979xh} which were found to describe lattice data satisfactory. Additionally, we consider next-to-leading order (NLO) expressions from Ref.~\cite{Tiburzi:2008bk}. The deviation of the values obtained at the physical pion mass from fitting to the leading-one loop expressions and to the NLO expressions provides an estimation of the systematic error due to the chiral extrapolation. Representative plots of the chiral fits on the octet and decuplet baryons are shown in the top panel of \fig{fig:chiral}. We note here that the results shown are continuum extrapolated and thus exhibit larger errors than those of the raw data. The error bands for all the fits were constructed using the so-called super-jackknife procedure~\cite{Bratt:2010jn}. In general, both fits describe the data satisfactory, though the NLO fits in general extrapolate to a lower value than that from the LO fits at the physical point.

In the charm sector we use the Ansatz $m_B = m_B^{(0)} +c_1 m_\pi^2 +c_2 m_\pi^3$, motivated by SU(2) HB$\chi$PT to leading one-loop order, where $m_B^{(0)}$ and $c_i$ are treated as independent fit parameters. In order to estimate a systematic error due to the chiral extrapolation in this case, we perform a linear fit w.r.t. $m_\pi^2$ by setting $c_2=0$, also restricting our lattice data only up to $m_\pi\sim 300$~MeV. The deviation of the values obtained at the physical pion mass from fitting using the whole pion mass range and fitting up to $m_\pi\sim 300$~MeV yields an estimation of the systematic error due to the chiral extrapolation. In the bottom panel of \fig{fig:chiral} we show representative fits on the $\Xi_c^0$ and $\Sigma_c^*$ baryons. As in the strange sector, our continuum extrapolated data are well described by this Ansatz. It is also apparent that setting $c_2=0$ in the Ansatz and fitting to the whole pion mass range would have led to satisfactory fits as well. This in part is reflected by the large uncertainty of this parameter, allowing it to be compatible with zero in most cases. 

A systematic error due to the tuning is also estimated for all strange and charm baryons. To do this, we  evaluate the baryon masses when the strange and charm quark masses take the upper and lower bound allowed by the error in their tuned values. The deviation between this value and the one extracted using the leading order $\chi$PT expressions provides an estimate of the systematic error due to the tuning.

As already mentioned,  continuum extrapolation is
performed  by including a term $da^2$ in all fit expressions.
Extrapolating to the continuum limit also ensures that the small non zero mass differences  due to isospin breaking effects observed for the $\Sigma^{+,0,-}$, $\Xi^{0,-}$ and $\Xi_c^{+,0}$ baryons vanish. The fit parameters, the lattice data and the values extracted at the physical point for all baryons are found in Ref.~\cite{Alexandrou:2014sha}.
%


\section{Results III: Comparison}

Several collaborations have studied the baryon spectrum, using a number of different lattice actions. It would be interesting to compare the results obtained from this work with those from other collaborations as a function of the pion mass as well as at the physical point and at the continuum, where comparisons with experiment can also be made.

\begin{figure}[!ht]\vspace*{-0.2cm}
\begin{minipage}{5cm}
\includegraphics[width=\textwidth]{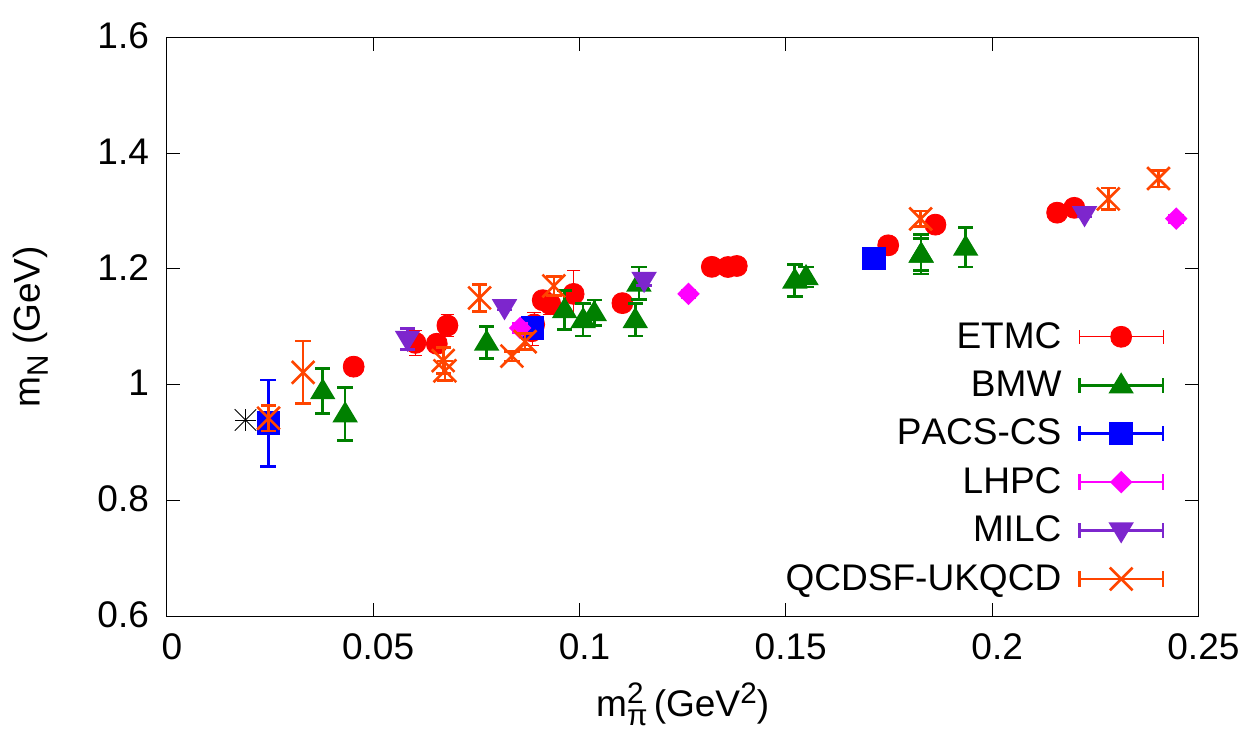}
\end{minipage}\hfill
\begin{minipage}{5cm}
\includegraphics[width=\textwidth]{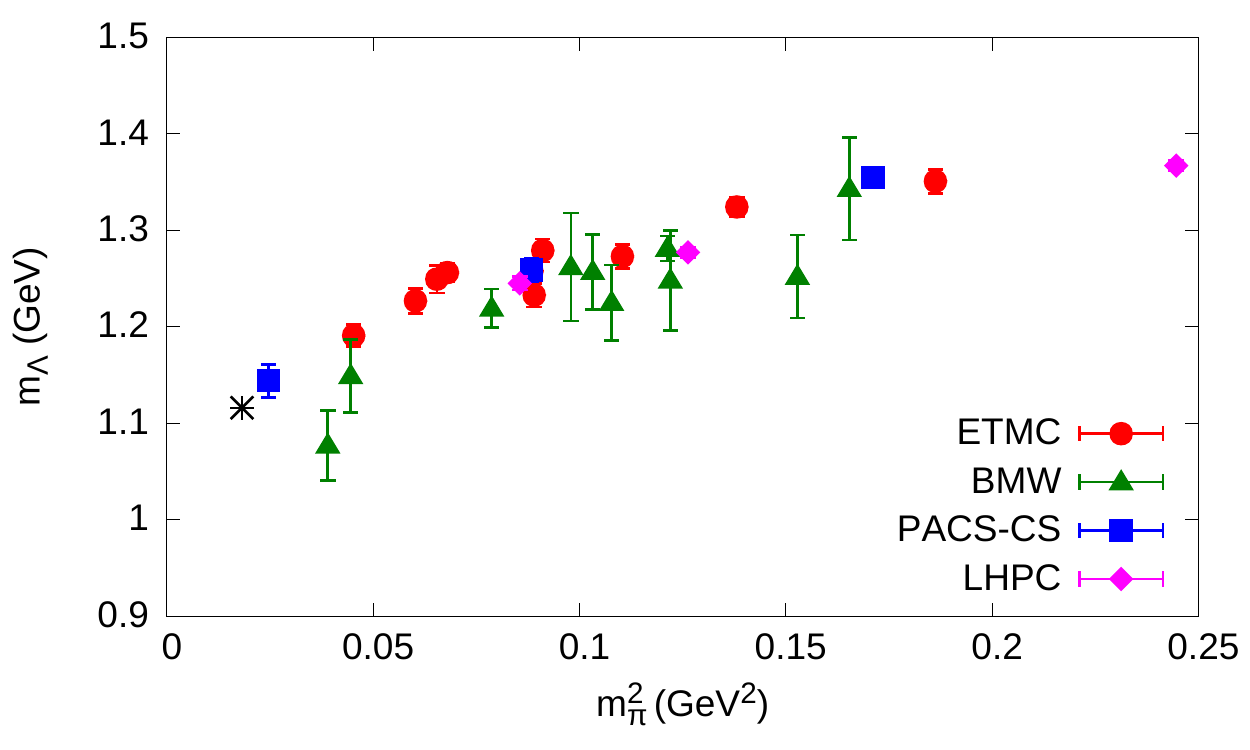}
\end{minipage}
\begin{minipage}{5cm}
\includegraphics[width=\textwidth]{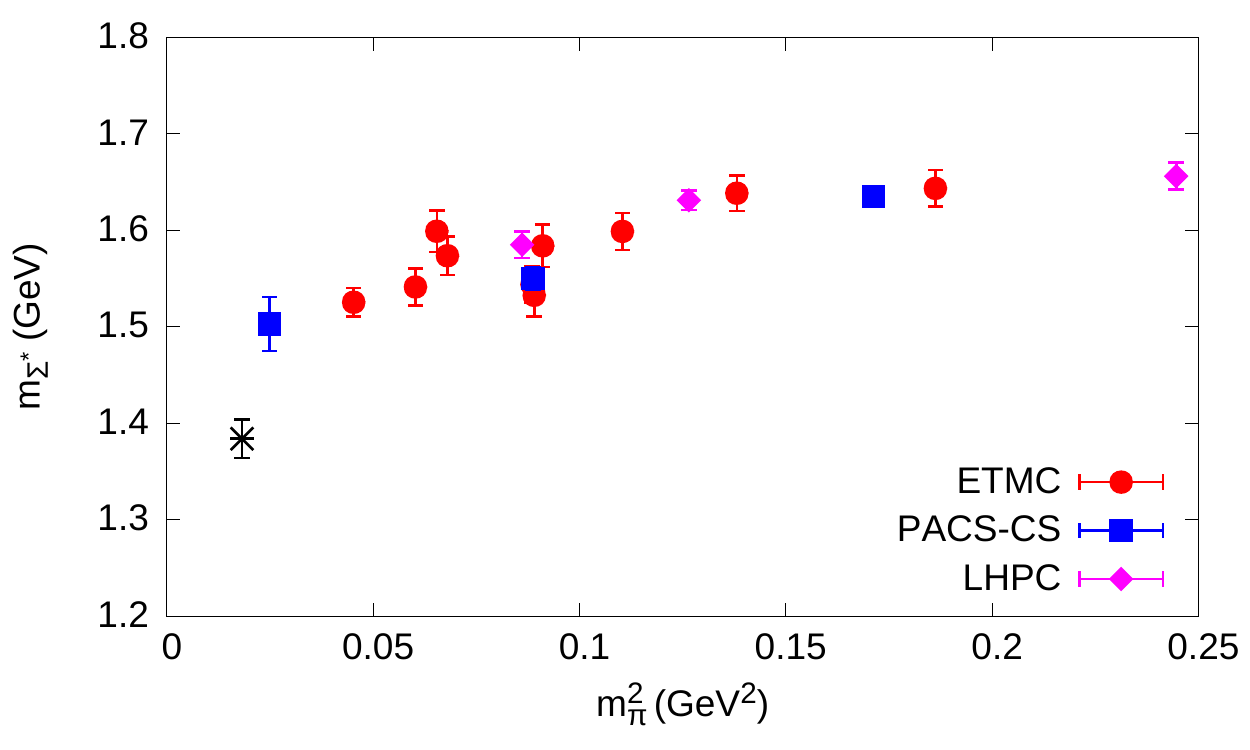}
\end{minipage}
\caption{Comparison of the lattice results of this work (ETMC) with those from other collaborations. From left to right: Nucleon mass, $\Lambda$ mass, $\Sigma^*$ mass. The notation is given in the legends of the plots.}
\label{fig:comparison}
\end{figure}

In \fig{fig:comparison} we show representative comparison plots of our lattice results on the octet and decuplet baryons with those from the BMW~\cite{Durr:2008zz}, the PACS-CS~\cite{Aoki:2008sm} and the LHPC~\cite{WalkerLoud:2008bp} collaborations. In the nucleon case, we furthermore compare with results from the MILC~\cite{Bernard:2001av} and QCDSF-UKQCD~\cite{Bali:2012qs} collaborations. As can be seen, there is an overall agreement which is best depicted in the nucleon mass, also indicating that cut-off effects are small. We note that at this point small deviations between different lattice actions are expected in the raw data since they still need to be continuum extrapolated to make more direct comparisons.
\begin{figure}[!ht]\vspace*{-0.2cm}
\begin{minipage}{5cm}
\includegraphics[width=\textwidth]{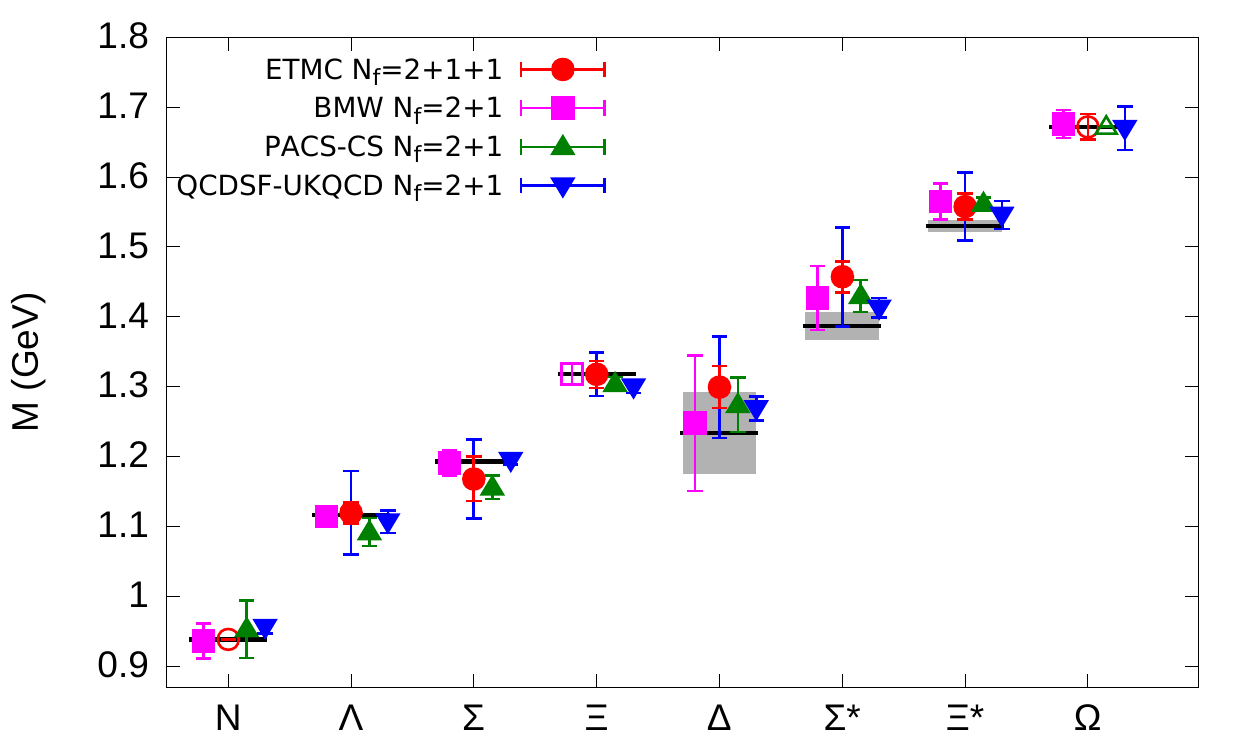}
\end{minipage}\hfill
\begin{minipage}{5cm}
\includegraphics[width=\textwidth]{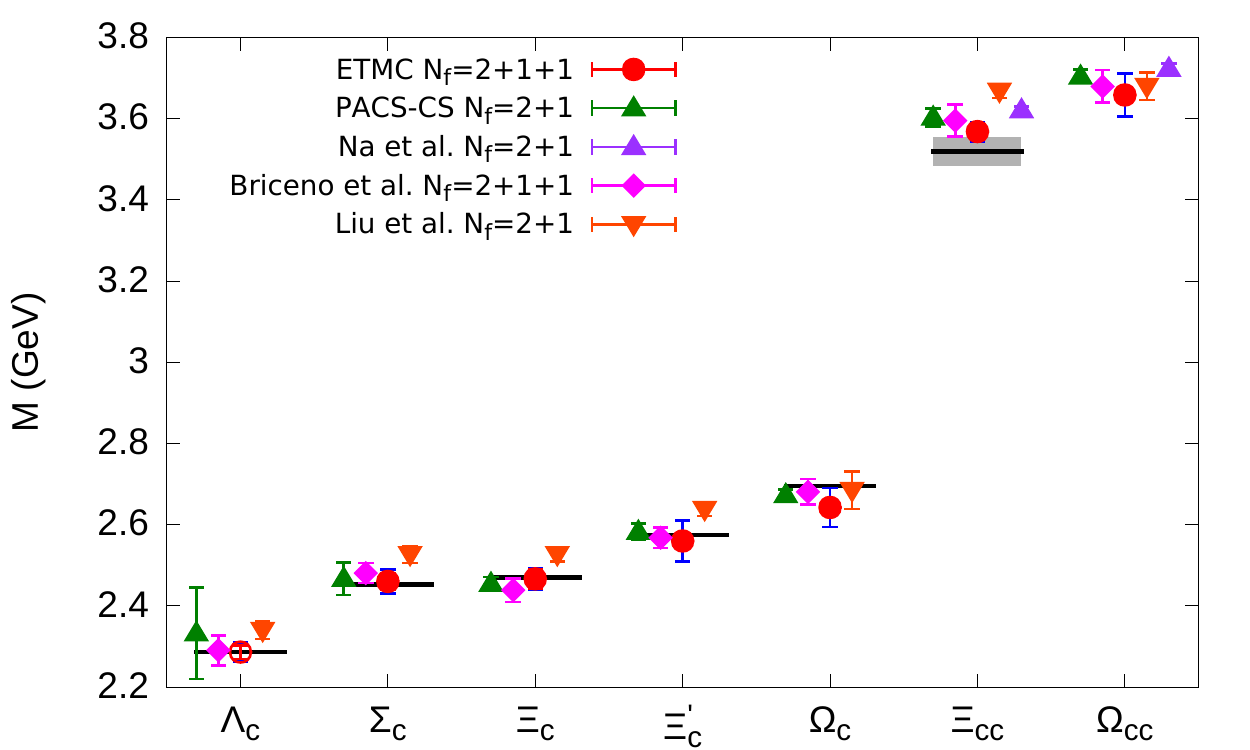}
\end{minipage}
\begin{minipage}{5cm}
\includegraphics[width=\textwidth]{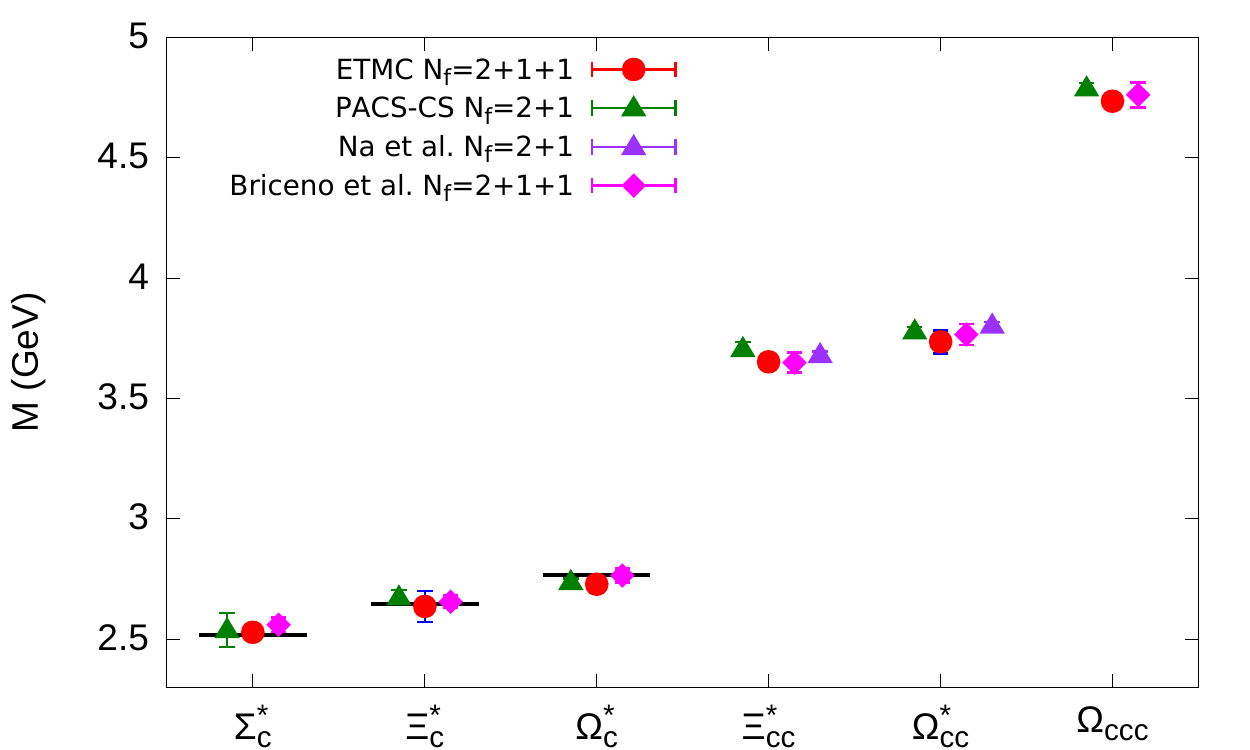}
\end{minipage}
\caption{The octet and decuplet baryon masses (left), the spin-1/2 charmed baryon masses (center) and the spin-3/2 baryon masses (right) as obtained in this work (ETMC) at the physical point. The experimental values are depicted by the horizontal bands. The notation of the results from other collaborations is shown in the legends of the plots.}
\label{fig:spectrum}
\end{figure}

In \fig{fig:spectrum} we show the octet and decuplet as well as the charmed baryon masses after extrapolating to the physical point as obtained in this work~\cite{Alexandrou:2014sha}. In these plots, the experimental values wherever available are also shown~\cite{Hagiwara:2002fs}, together with results from a number of other lattice calculations~\cite{Durr:2008zz,Aoki:2008sm,Bietenholz:2011qq,Briceno:2012wt,Na:2008hz,Liu:2009jc,Na:2007pv,Namekawa:2013vu}, as labeled in the legends of the plots.  In our results (ETMC), the statistical error is shown in red, whereas the blue error bar includes the statistical error and the systematic errors due to the chiral extrapolation and due to the tuning of the strange and/or charm quark mass added in quadrature. As can be seen, our results are consistent with the experimental values as well as with the results from the other collaborations. Our value for the $\Xi_{cc}$ is also within errors with the experimental one.
Given the agreement with experiment, our LQCD calculation provides predictions for the masses of doubly and triply charmed baryons that have not yet been measured experimentally.
The value we find for the mass of $\Xi_{cc}^*$ is 3.652(17)(27)(3) GeV, for $\Omega_{cc}$ is 3.658(11)(16)(50) GeV, for  $\Omega_{cc}^*$ is  3.735(13)(18)(43) GeV and for $\Omega_{ccc}$ is 4.734(12)(11)(9) GeV, where the error in the first parenthesis is the statistical, in the second the systematic due to the chiral extrapolation and in the third the systematic due to the tuning.


\section{Conclusions}

The twisted mass formulation allowing simulations with dynamical strange and charm quarks with their mass fixed to approximately their physical values provides a good framework for studying the baryon spectrum.  When extrapolating our results to the physical pion mas and the continuum, we find that the largest systematic uncertainty arises from the chiral extrapolation. Our results are compatible
to those of  other lattice calculations. After extrapolating to the physical pion mass and the continuum, we find remarkable agreement with experiment, which allows for reliable predictions for the mass of $\Xi_{cc}^*$, $\Omega_{cc}$, $\Omega_{cc}^*$ and $\Omega_{ccc}$.

\vspace{0.4cm}
{\bf Acknowledgments:} The project used computer time at JSC granted by the John von Neumann Institute for Computing (NIC) and at the Cy-Tera machine under the Cy-Tera project (NEA $\Upsilon\Pi$O$\Delta$OMH/$\Sigma$TPATH/0308/31). C. Kallidonis is supported by the project GPUCW (T$\Pi$E/$\Pi\Lambda$HPO/ 0311(BIE)/09).

\bibliographystyle{./apsrev}
\tiny
\bibliography{refs}

\end{document}